==================================================================

                        Re-submission information

Journal:                Journal of Physics A: Math. Gen.

Your ref.:              A/102400/PAP

Article title:          Harmonic oscillator well with a screened
                        Coulombic core is quasi-exactly solvable

Author:                 Miloslav Znojil

Article Type:           Paper

Status of article:      revised

Reference number:       A/102400/PAP

Postal address:         Miloslav Znojil
                        Nuclear Physics Institute
                        250 68 \v{R}e\v{z},
                        Czech Republic

E-mail address:         znojil@ujf.cas.cz

Phone number:           (00420-2) 6617 3286

Fax number:             (00420-2) 6857003

Article file format:     Latex

Number of figures:       0

Enclosures:              2 replies to the referees about changes
                         made

==================================================================

Encl. 1: Reply to the first referee

1) Section 1.2 extended and moved to the end of the text (where it
originally was - my proof of QES precedes it in fact). More
details added, misprints corrected. Thanks for you comments.

Marginally: You are right: The addition of parameter c is
responsible for the emergence of $D_{n}$. The change of sign
reflects the fact that the Lie algebras in question are only
equivalent over the complex field. In fact, up to an inessential
shift of $y$, my algebra coincides with the one of ref. [9].

2) Thanks for your having attracted my attention to the necessity
of an explicit comment on boundary conditions after the change of
variables. Clarified. Discussion added.

3) The novelties of perturbative approach etc re-emphasized. More
comments on non-hermiticity added.

Marginally: You are right: This is an old business. Still, I've
met the two-diagonal $H^{[0]}$ here with pleasure.

4) Thanks for the detected misprints.

==================================================================

Encl. 2: Reply to the second referee:

1. Abstract made more accurate (Note: E is fixed!). Here and
there, a few sentences all over the text were re-formulated.

2. Several typogr. err. corrected (especially in the orig. last
eq. on p. 2 and the first one on p. 3, plus accompanying text).

3. Section 1.2 moved from Introduction to the final Discussion,
with a few more comments added, esp. on [9] and other references.
It would be nice to know a transformation!

Thanks for your advice.

========================Latex file==========================================

\documentstyle[12pt]{article}

\def\be{\begin{equation}}
\def\ee{\end{equation}}
\def\ba{\begin{array}{c}}
\def\ea{\end{array}}

\def\ben{$$}
\def\een{$$}
\begin{document}

\titlepage
%\vspace*{2cm}

\begin{center}{\Large \bf
Harmonic oscillator well
 with a screened Coulombic core
 is
quasi-exactly solvable
 }\end{center}

\vspace{5mm}

\begin{center}
Miloslav Znojil
\vspace{3mm}

\'{U}stav jadern\'e fyziky AV \v{C}R, 250 68 \v{R}e\v{z},
Czech Republic\\

\end{center}

\vspace{5mm}

\section*{Abstract}

In the quantization scheme which weakens the hermiticity of a
Hamiltonian to its mere ${\cal PT}$ invariance the superposition
$V(x) = x^2+ Ze^2/x$ of the harmonic and Coulomb potentials is
defined at the purely imaginary effective charges ($Ze^2=if$) and
regularized by a purely imaginary shift of $x$. This model is
quasi-exactly solvable: We show that at each excited, $(N+1)-$st
harmonic-oscillator energy $E=2N+3$ there exists not only the well
known harmonic oscillator bound state (at the vanishing charge
$f=0$) but also a normalizable $(N+1)-$plet of the further
elementary Sturmian eigenstates $\psi_{\{n\}}(x)$ at eigencharges
$f=f_{\{n\}}>0$, $n = 0, 1, \ldots, N$. Beyond the smallest
multiplicities $N$ we recommend their perturbative construction.

\vspace{9mm}

\noindent
 PACS 03.65.Ge,
03.65.Fd

%\vspace{9mm}

\begin{center}
%{\small \today, chart.tex file}
\end{center}

\newpage

\section{Introduction}

Schr\"{o}dinger equation for all the asymptotically harmonic
oscillators in one dimension,
 \be
\left [-\,\frac{d^2}{dx^2} + x^2+2\alpha\,x -E + {\cal O}
 (1/x)
\right ] \, \psi(x) = 0,
\label{one}
 \ee
need not necessarily be kept defined just on the real axis of
coordinates. Indeed, its available physical asymptotic solutions
 \be
\psi (x) \sim
 \exp\left[-\frac{x^2}{2}- \alpha\, x
+ b\, {\rm ln} \,(x) + {\cal O} (1/x)
 \right]\ ,\ \ \ \ \
 b =(E+\alpha^2-1)/2
 \label{polansatz}
 \ee
are normalizable not only on the real intervals $x > x_a\gg 1$ and $x <
-x_a$ but also in their complex vicinity defined by the respective
formulae
 \be |x| \in (x_a, \infty), \ \ \ \ {\rm arg}\ x \in
(-\pi/4, \pi/4)\ \ {\rm and}\ \  {\rm arg}\ x \in (-5\pi/4,
-3\pi/4). \label{wedges}
 \ee
One may connect these two complex wedges by a contour of
integration which is arbitrarily deformed within the domain of
analyticity of the potential in question.  In particular, one may
choose the purely harmonic $V(x)=x^2$ and move the real axis of
coordinates $x$ to a parallel line at a distance $a$. The new,
shifted potential $V(x) = x^2 + 2aix - a^2$ loses its hermiticity
and preserves only a certain symmetry with respect to a
simultaneous change of the parity ${\cal P}$ ($x\to -x$) and of
the time ordering ${\cal T}$ (which means the mere complex
conjugation $i \to -i$ in time-independent cases), {\em without
any change in the discrete spectrum itself}.

Similar paradoxes seem to have inspired a deeper analysis of the
various ${\cal PT}$ symmetric phenomenological models. Bessis and
Zinn-Justin \cite{Bessis} related some of them to the so called
Lee-Yang zeros in field theory \cite{Itzykson} and Bender with
coauthors paid attention to their possible role in the parity
breaking \cite{BM}, phase transitions \cite{BBprl} and quantum
electrodynamics \cite{BMb}.

${\cal PT}$ symmetry does not prove less exciting on a purely
methodical level. Its use may range from the fundamental problems
of the ambiguities of quantization \cite{Meis} and of the exact
solvability and Darboux transformations \cite{Cannata} up to the
questions of convergence of perturbation expansions
\cite{myctyri}. In this context, there also appeared an amazing
discovery \cite{BBjpa} of the so called quasi-exact (which means
incomplete \cite{Ushveridze}) solvability of quartic oscillators.

In contrast to its unsolvable three-dimensional counterpart the
latter quartic model did not contain the Coulombic component $
e^2/x $ \cite{Classif}.  In all the one-dimensional Hermitean
models this term is routinely being omitted due to its strongly
singular character in the origin. In the ${\cal PT}-$symmetric
non-hermitean setting, nevertheless, a purely imaginary shift {\em
could} regularize this singular term in principle. This was our
main inspiration.

Even after the screening $x \to x-ic$  the persistent imaginary
part of forces $V(x) \sim 1/(x-i\,c)$ would still violate
unitarity and cause the predominance of unstable, resonant bound
states.  At this point we may recall the above-mentioned
strategy which tries to compensate the instabilities by the
constraint of ${\cal PT}$ symmetry.  This re-defines the
electric charge. With real $f$ in its effective ${\cal
PT}-$symmetric value $Z e^2\equiv if$ the nontrivial interaction
model need not even contain the cubic and quartic terms. Its
Schr\"{o}dinger equation with the three real parameters $a$, $f$
and $c$ reads
\be
\left[-\,\frac{d^2}{dx^2} + x^2+2iax + i\frac{f}{x-i\,c}\right]\,
\psi(x) = E \psi(x), \ \ \ \ \  \psi(\pm \infty) = 0,
 \label{SE}
 \ee
and gives, presumably, a real and discrete spectrum of energies
$E$.

In what follows, we shall re-examine eq. (\ref{SE}) from the point
of view of its possible quasi-exact solvability. We shall be able
to show that its elementary $(N+1)-$plets of Sturmian eigenstates
exist at any $N=0, 1, \ldots$ (Section 2) and that their explicit
construction (mediated by the vanishing of an $(N+1) \times
(N+1)-$dimensional secular determinant, see below) may be
significantly facilitated via perturbative techniques (Section 3).
A summary of our results will be outlined in Section 4.

\section{Quasi-exact solutions}

\subsection{Taylor series and its termination}

Equations (\ref{one}) and (\ref{polansatz}) clarify the structure
of all the normalizable solutions $\psi(x)$ of eq. (\ref{SE}) near
$x = \pm \infty$. In the light of the identity $1/(x-ic) \equiv
(x+ic)/(x^2+c^2)$ our model remains smooth and regular at all the
finite coordinates and everywhere off the complex pole of $V(x)$
at $x = ic$. In the vicinity of this complex point it is
convenient to demand that our wave functions vanish, $\psi (x)
\approx x-ic$. Having in mind the possible deformations of the
contour of integration, such a requirement is immediately inspired
by the universal and widely accepted regularization of
Schr\"{o}dinger equations near their strong singularities
\cite{Frank}. On this basis let us now try to solve our present
Schr\"{o}dinger bound state problem (\ref{SE}) by means of the
following elementary harmonic-oscillator-like ansatz
\be
\psi (x) = (c + ix)\,
 \exp\left[-\frac{x^2}{2}- iax\right]\,
\varphi(x), \ \ \ \ \ \ \ \ \ \varphi(x) =
 \sum_{n=0}^{N}
\,h_n \,(ix)^n.  \label{ansatz}
 \ee
Such a terminating Taylor-series assumption fixes the energy (cf.
eq. (\ref{polansatz})),
 \be
E=2N+a^2+3. \label{energy}
 \ee
Its insertion in our differential Schr\"{o}dinger bound state
problem (\ref{SE}) leads to the equivalent $N+1$ recurrence
relations
 \ben
A_nh_{n-1}+B_nh_{n}+C_n h_{n+1} + D_n h_{n+2}=0, \ \ \ \ \
 \ \ \ \ \ \ \ \ n = 0, 1, \ldots, N.
 \een
Keeping in mind that $h_{-1} = h_{N+1}=h_{N+2}=0$ we easily
derive the values of the coefficients,
 \ben D_n=c(n+1)(n+2), \ \ \ \
C_n=(n+1)(n+2-2ac), \een
 \ben B_n =
-2a(n+1) - 2c(N+1-n)-f, \ \ \ \ \ \  A_n = -2(N+1-n).
 \een
They may be arranged in a square matrix with four diagonals,
 \ben Q  =
 \left( \begin{array}{cccccc}
 B_{0} & C_0& D_0 & & & \\
A_1&B_{1} & C_1& D_1   && \\ & A_2&\ddots & \ddots& \ddots   & \\
&&\ddots&\ddots&\ddots&D_{N-2}\\ &&&A_{N-1}&B_{N-1}&C_{N-1}\\
&&&&A_N&B_N
\end{array} \right).
 \een
In notation with the row vectors $\vec{h}= (h_0, h_1, \ldots,
h_N)$ our recurrences may be then re-interpreted as a
non-hermitean matrix problem
 \ben
  Q\,\vec{h}^T=0.
  \een
Normalization $h_N=1$ implies its immediate compact and unique
solution
 \be h_{N-k-1}=\frac{1}{2^{k+1}\,(k+1)!} \det \left(
\begin{array}{ccccc} B_{N-k} & C_{N-k}& D_{N-k} &  & \\
A_{N-k+1}&\ddots & \ddots& \ddots &
\\ &\ddots&\ddots&\ddots&D_{N-2}\\ &&A_{N-1}&B_{N-1}&C_{N-1}\\
&&&A_N&B_N
\end{array} \right)
\label{defi}
 \ee
for all the relevant indices $k = 0, 1, \ldots, N-1$.  At the
remaining, ``redundant" value of $k = N$ the left-hand-side symbol
vanishes identically since, due to our assumptions, $h_{-1} \equiv
0$. With the linear charge-dependence in $B_n=B_n(f)=B_n(0)-f$ and
in $Q = Q(f)=Q(0)-fI$ this is equivalent to the current secular
equation
 \be
  \det\left[ Q(0) - f I\right] =
0. \label{seculare}
 \ee
As a polynomial constraint of the $(N+1)-$st degree this
determines an $(N+1)-$plet of eigencouplings $f=f_{\{k\}}$. They
are complex in general. In the spirit of our introductory remarks
it only remains for us to demonstrate that all these roots are
real (i.e., not breaking the overall ${\cal PT}-$symmetry of our
model), in a certain sub-domain of parameters at least.

\subsection{Full ${\cal PT}$ symmetry in the large-screening domain}

For convenience let us shift the variable $f = X+(N+2)(c-a)$ and
look at Table~1. It lists the explicit secular equations at the
first few lowest integers $N$. From the Table we may infer that
all the zeros $X_{\{n\}}$ and/or $f_{\{n\}}$ of eq.
(\ref{seculare}) are just functions of the single parameter $d =
c+a$. This is not surprising. The change of the value of $a$ is
just a shift of the integration path within the domain of
analyticity of our potential. We may demonstrate the explicit
form of this type of invariance of our differential eq.
(\ref{SE}) algebraically: The equation remains the same after we
compensate the change of the coordinate $x \to x+i\delta$ by the
simultaneous shift $a \to a+\delta$ and $c \to c - \delta$ of
our pair of parameters.  The values of $f$ and $d$ remain
unchanged. In our subsequent considerations we shall put
$a=0$ without any loss of generality, therefore.

Assuming that the strong Coulomb singularity lies off the real
line, $c \neq 0$, we may introduce $\lambda = 1/c$ and re-scale $Y
= X/c$. Our secular equation (\ref{seculare}) may be re-phrased as
corresponding to the asymmetric linear algebraic problem
\be
 \left( \begin{array}{cccccc}
 -Y-N & 2\lambda& 2 & & & \\
-2N\lambda&-Y-N+2 & 6\lambda& 6   && \\ & (2-2N)\lambda&\ddots &
\ddots& \ddots &
\\ &&\ddots&\ddots&\ddots&N(N-1)\\ &&&-4\lambda&-Y+N-2&N(N+1)\lambda\\
&&&&-2\lambda&-Y+N
\end{array} \right) \left( \begin{array}{c} h_0\\ h_1\\ h_2\\
\vdots\\ h_{N-1}\\ h_N
\end{array} \right)
 = 0
. \label{linear}
 \ee
In the large-displacement limit $\lambda \to 0$ we get an exactly
solvable case which determines the real eigencharges $f_{\{n\}}$.
This may be proved easily since in this limit our ``Hamiltonian"
$\lambda Q = H- Y\,I$ becomes an upper triangular matrix. This
leads to the following closed formula
 \ben \left (Y_{\{0\}}, Y_{\{1\}},
\ldots, Y_{\{N\}}
 \right )   \to
\left (Y^{[0]}_{\{0\}}, Y^{[0]}_{\{1\}}, \ldots, Y^{[0]}_{\{N\}}
 \right )   =
 \left (N, N-2, \ldots ,
-N+2, -N\right )
 \een
which determines the real and approximately equidistant spectrum
of eigencharges $f=f_{\{n\}} (c)= 2(n+1)c+{\cal O}(1/c)$ with $n =
0, 1,\ldots, N$.

\section{Sturmian bound states at the general $N$}

Once we fix the integer $N$ and recall the nonlinear algebraic
definition (\ref{seculare}) of our $(N+1)$ eigencouplings $f =
f_{\{n\}}$ we immediately imagine that the explicit construction
of our Sturmians becomes more and more complicated for the larger
dimensions $N$. Even if we extend, accordingly, the list of our
secular polynomials in Table 1 we must necessarily resort to the
purely numerical methods when trying to determine their exact
roots $f$. The task is easy for the first few integers $N$ only.

At the large $N$, our numerical algorithms may still start from
the above equidistant asymptotic estimates and make use of the
expected smooth change of the roots $f$ with the decrease of the
parameter $|d|<\infty$. Indeed, the boundary of the domain $d>
d_{critical}(N)$ where all our roots $f$ stay real is only slowly
growing with $N$ since $d_{critical}(1)=2$,
$d_{critical}(2)\approx 2.9865$, $d_{critical}(3)\approx 3.765$
etc.

All this indicates that in practice a perturbative evaluation of
the eigencharges could prove almost as efficient as their direct
numerical determination, especially beyond the smallest $N$.

\subsection{Perturbation expansions with $\lambda<\lambda_{critical}$}

The inspection of the (extended) Table 1 reveals that
 \be  Y(\lambda)=Y^{[0]} + \lambda^2\,Y^{[2]} + \lambda^4
 Y^{[4]} + \ldots\ .
 \label{eners}
 \ee
This means that all the odd perturbation corrections vanish
identically, $Y^{[2k+1]}=0$. In the similar spirit we may also
write
  \be
\vec{h}=\vec{h}(\lambda) = \vec{h}^{[0]} + \lambda\,\vec{h}^{[1]}
+ \lambda^2 \vec{h}^{[2]} + \ldots\ . \label{coeff}
 \ee
In order to appreciate the possible merits of such an approach let
us return to our asymmetric eq. (\ref{linear}) and notice that at
$\lambda=0$ it decays into a direct sum of the two linear
equations. Each of them couples only $h_n$'s with the same parity
of the subscript $n$. This is a pleasant simplification. For
example, the five-dimensional unperturbed eigenvalue problem at
$N=4$ decays into the separate two- and three-dimensional
sub-equations. We may omit the redundant superscripts~$^{[0]}$ and
display the latter subset for illustration,
\be
 \left( \begin{array}{ccc}
 -Y-4 & 2& 0 \\
0 & -Y & 12\\ 0& 0& -Y+4
\end{array} \right)
 \left( \begin{array}{c}
h_0\\ h_2\\ h_4
\end{array} \right)
 = 0.
 \label{trinear}
 \ee
Relaxing our above normalization convention and working in the
integer arithmetics (i.e., in full precision, without any
round-off errors) its first solution $ \vec{h}_{\{a\}}=(1,0,0)$ is
found for the eigencharge $Y=Y_{\{a\}}=-4$ while
$\vec{h}_{\{b\}}=(1,2,0)$ is obtained at the vanishing
$Y_{\{b\}}=0 $ and $\vec{h}_{\{c\}}=(3,12,4)$ results from
$Y_{\{c\}}=+4$.

Returning to our matrix form $ Q\,\vec{h}^T=0$ of the differential
Schr\"{o}dinger eq. (\ref{SE}) with ``Hamiltonian" $\lambda Q = H-
Y\,I$ and decomposition $H=H^{[0]}+\lambda\,H^{[1]}$ at a general
$N$ we may now solve it by means of the textbook
Rayleigh-Schr\"{o}dinger perturbation theory \cite{Messiah}. In
the present implementation of this recipe the $k-$th unknown
perturbation corrections will be determined by the relation
 \ben
 \left (H^{[0]}-Y^{[0]}I\right )
 \left (
 \vec{h}^{[k]}
 \right )^T
+
 \left (H^{[1]}-Y^{[1]}I\right )
  \left (
  \vec{h}^{[k-1]} \right )^T -\ \ \ \ \ \ \ \ \ \ \ \ \ \ \ \ \
  \een
  \ben
  \ \ \ \ \ \ \ \ \ \ \ \ \ \ \ \ \ \ \
 -Y^{[2]}
  \left (
  \vec{h}^{[k-2]} \right )^T - \ldots  -Y^{[k]}
   \left (
   \vec{h}^{[0]} \right )^T=0.
 \label{RSE}
  \een
We have to solve it step-by-step, at all the subsequent
perturbation orders ${\cal O}(\lambda^k) $ numbered by the integer
$k=1, 2, \ldots$. For illustration let us pick up $N=2$ and $
Y^{[0]}=2$.  In the first-order approximation with $k=1$ we
normalize $h_N^{[1]}=0$ and drop all the superscripts $^{[1]}$
from $Y^{[1]}$ and $\vec{h}^{[1]}$. The perturbed equation
(\ref{linear}) then acquires its ${\cal O}(\lambda)$ first-order
form
 \ben
 \left( \begin{array}{rrr} -4 & 0&+ 2 \\ 0 & -2 & 0\\ 0& 0& 0
\end{array} \right)
 \left( \begin{array}{c} h_0\\ h_1\\ 0
\end{array} \right)
+ \left( \begin{array}{rrr} -Y & 2& 0 \\ -4 & -Y & 6\\ 0& -2& -Y
\end{array} \right)
 \left( \begin{array}{c} 1\\ 0\\ 2
\end{array} \right) = 0.
 \label{twonear}
  \een
This implies that $Y = Y^{[1]}=0$ (as expected) and
$\vec{h}=\vec{h}^{[1]}=(0,4,0)$. By the way, the latter result
nicely illustrates a general feature: We might replace eq.
(\ref{coeff}) by a component-by-component expansion in the powers
of squares $\lambda^2$ in a way paralleling the expansion of
charges $Y$ above. This reflects a symmetry of our model with
respect to the change of sign of the parameter of screening $d$.

\subsection{The role of asymmetry}

We have established that our recurrences (\ref{linear}) provide a
new, not entirely standard framework for application of
perturbation theory. The manifestly non-hermitean two-diagonal
structure of the underlying exactly solvable unperturbed
Hamiltonians $H^{[0]}$ requires a modification of the formalism
itself. The necessity of re-considering the standard concepts
(say, of the model-space projector) seems to deserve a separate
comment. The point is that due to the manifest difference between
our original and transposed pseudo-Hamiltonians we must consider
not only the ``direct" zero-order eigenvalue problem $ [Q(0)-
f\,I] \left (\vec{h}^{[0]}\right )^T = 0$ (or rather equation
 \ben
\left (H^{[0]} -Y_{\{\alpha\}}\,I\right ) \left (
\vec{h}_{\{\alpha\}}^{[0]}\right )^T= 0
 \een
at a fixed subscript $\alpha = 0, 1, \ldots N$) but also its
transposed, non-equivalent pendant
 \ben
\left [ \left (H^{[0]}\right )^T-Y_{\{\alpha\}}\, I \right]\left (
\vec{g}_{\{\alpha\}}^{[0]}\right )^T  = 0.
 \een
It is worth noticing that in a way paralleling eq. (\ref{defi})
above, all the left eigenvectors $\vec{g}=\vec{g}(\lambda)$ may
be defined by closed formulae,
 \ben
g_{k+1}=\frac{(N-k-1)!}{2^{k+1}\,(k+1)!} \det \left(
\begin{array}{ccccc} B_{0} & C_{0}& D_{0} &  & \\ A_{1}&\ddots &
\ddots& \ddots &
\\ &\ddots&\ddots&\ddots&D_{k-2}\\ &&A_{k-1}&B_{k-1}&C_{k-1}\\
&&&A_k&B_k
\end{array} \right) .
\label{defipr}
 \een
Their knowledge would simplify the large-order algorithm. Still,
due to the unphysical, auxiliary character of all the left
eigenvectors, it is shorter to generate and use them just in the
zero order. Dropping their superscript $^{[0]}$ as redundant, we
may return to our illustration (\ref{trinear}) and complement it
by the left eigenvectors $\vec{g}_{\{a\}} =(4,-2,3)$,
$\vec{g}_{\{b\}}=(0,1,-3)$ and $\vec{g}_{\{c\}} =(0,0,1)$.
Incidentally, the overlaps $G = (g,h)$ form a diagonal matrix,
$G_{11}=G_{33}=4$, $G_{22}=2$. This trivializes $F = G^{-1}$
needed in the model-space projectors
 $
 P_{\{\alpha\}}=
\vec{h}_{\{\alpha\}}^T F_{\alpha, \alpha} \vec{g}_{\{\alpha\}}.
 $
They are, counter-intuitively, non-diagonal but
compatible with the usual completeness relation
 $
  I =
\sum_{\alpha,\beta}\, \vec{h}_{\{\alpha\}}^T F_{\alpha, \beta}
\vec{g}_{\{\beta\}}
 $.

\section{Concluding remarks}

\subsection{Sturmians and the algebra $sl(2)$.}

Our wave function (\ref{ansatz}) becomes purely real after the
change of variables $x = -iy$ which rotates the axes by $\pi/2$.
Of course, the exponential growth of our asymptotics
(\ref{polansatz}) in the new artificial variable $y$ has no
meaning at all and only the original coordinate $x$ remains
physical. Still, the use of $y$ simplifies our ansatz
(\ref{ansatz}). With $a=0$ it leads to a new differential
Schr\"{o}{dinger equation $\hat{H}\hat{\varphi}(y) =
f\,\hat{\varphi}(y)$ for real polynomials $\hat{\varphi}(y)=
\varphi(-iy)$ themselves. In this notation our third
Hamiltonian-like operator $\hat{H}$ is the quadratic function of
generators of the complex $sl(2)$ Lie algebra,
 \ben \hat{H}={\cal J}^0
{\cal J}^{-} +2
 {\cal J}^{+}-2c
  {\cal J}^{0} + (N+2) {\cal J}^{-}
-2(N+c)
   \een
   with
 \ben
\left [ {\cal J}^{-} {\cal J}^{0}\right ] =  {\cal J}^{-} =
\frac{d}{dy},
 \een
 \ben
\left [ {\cal J}^{-} {\cal J}^{+}\right ] = 2 {\cal J}^{0} =
2(y+c) \frac{d}{dy} -2N,
 \een
 \ben
\left [ {\cal J}^{0} {\cal J}^{+}\right ] =  {\cal J}^{+} =
(y+c)^2 \frac{d}{dy} -2(y+c)N.
 \een
This fits the general scheme \cite{Turbiner}, parallels closely
the similar quartic-oscillator result of ref. \cite{BBjpa} (with
the same algebra but different $\hat{H}$) and re-confirms our
above conclusion that in spite of its non-hermiticity, our ${\cal
PT}-$symmetric Schr\"{o}dinger eq. (\ref{SE}) is quasi-exactly
solvable.

In this context, we would like to emphasize that there exists a
nice parallel between the Hermitean and ${\cal PT}-$symmetric
quasi-exactly solvable models. Picking up the characteristic
examples and adding our present results, we may now summarize and
list the following four different possibilities and types of the
quasi-exact constructions.

\begin{itemize}

\item
The characteristic polynomial example $V(x) =
\alpha\,x^2+\beta\,x^4+x^6$ with a constrained variability, say,
of the coupling $\alpha$ leads to the existence of some $N$
elementary bound states at a finite multiplet of the binding
energies in the hermitean case \cite{Singh}.

\item
With a constrained variability of the energy $E$ the
non-polynomial hermitean potentials exemplified by $V(r) =
x^2+\alpha/(1+\beta\,x^2)$ lead to the elementary Sturmian
solutions at a finite $N-$plet of couplings $\alpha$
\cite{Flessas}.

\item
In the ${\cal PT}-$symmetric case, the complex quartic polynomial
potential of ref. \cite{BBjpa}, $V(x) = \alpha\,x+\beta\,
x^2+\gamma\,x^3-x^4$ with a constrained variability of $\alpha$,
exhibits the quasi-exact solvability of the sextic-oscillator
type.

\item
$V(x) = x^2 +i\alpha\,x+i\beta\,(x+i\gamma)/(x^2+\gamma^2)$ of the
present model (\ref{SE}) is the ``missing" ${\cal PT}-$symmetric
partner to the quasi-exact solvability of the non-polynomial
Sturmian type.

\end{itemize}

\subsection{Complex charges}

We have seen that the one-dimensional Schr\"{o}dinger
equation becomes quasi-exactly solvable for Coulomb plus harmonic
superpositions of potentials provided only that we regularize
these forces in the ${\cal PT}-$symmetric manner. Multiplets of
the Sturmian eigenstates acquire then an elementary polynomial
form at certain purely imaginary couplings $Z e^2 = i\,f_{\{n\}}$
at $n = 0,1, \ldots, N$ and any $N = 0, 1, \ldots$.

In a tentative physical support of these {\em complex} electric
charges we might recall a few of their ${\cal PT}$ symmetric
predecessors. A close connection exists with the Bessis' cubic
force possessing a purely imaginary coupling $g = i\,f$. Its
rigorous mathematical analysis has already been delivered, many
years ago, by Calicetti et al \cite{Calicetti}. One may also
mention an even closer parallel with the Bender's and Milton's
electrodynamics which replaces, for several independent reasons
\cite{BMb}, the charge $e$ itself (i.e., not its present square
$e^2$) by a purely imaginary quantity.

In a broader methodical context, our new phenomenological model
(\ref{SE}) extends the family of potentials which are
comparatively easily described in the language of analytic
continuations \cite{Alvarez}. Its distinctive feature, in this
setting, is the presence of a complex pole.  Its most important
formal merit is its quasi-exact solvability shared with the
quartic model of ref. \cite{BBjpa}. In the latter comparison, one
could emphasize the ``more natural" asymptotic behaviour of our
present wave functions: The real axis of coordinates still lies
within the wedges (\ref{wedges}) of the admissible analytic
continuations.

In the context of perturbative considerations it is amusing to
notice that a ``nice" (which means real, discrete and bounded)
character of spectrum of our present ${\cal PT}$ symmetric model
may be viewed as a result of perturbation of {\em any} one of its
{\em two} exactly solvable halves.

\subsection*{Acknowledgement}
Partly carried out within the frame of TMR - Network PECNO -
ERB-FMRX-CT96-0057.

\vspace{1cm}

%\newpage

Table 1. The first six secular equations (\ref{seculare}) for
eigencharges $f + (N+2)(a-c) \equiv X$, with a variable parameter
$c+a\equiv d$ and abbreviation $d^2-N-3 \equiv h$.

$$
 \begin{array}{|c|c|} \hline
N & \det Q(f)=0\\ \hline
 & \\
0&-X=0\\ 1&{X}^{2}-h=0
\\
2&-{X}^{3}+4\,h{X}+8\,d=0
\\
3&{X}^{4}-10\,h{X}^{2}-48\,d X +9\,h^2-36 =0
\\
4&-{X}^{5}+20\,h{X}^{3}+168\,d {X}^{2}-32\,\left (2\,h^2-9\right )
X-384\,h d=0
\\
5&{X}^{6}-35\,h{X}^{4}-448\,d{X}^{3}+\left (
259\,{h}^{2}-1296\right ){X}^{2}+ \ \ \ \ \ \ \ \ \ \ \ \ \ \
\\ & \ \ \ \ \ \ \ \ \ \ \
\ \ \ \ \ \ \ \ \ \ \ \ \ \ \ \ \ \ \ \ \ + 3520\,h d X
-225\,{h}^{3}+10000\,h+51200=0\\ &
\\ \hline
\end{array}
$$

\end{document}